\documentclass{mn2e}
\usepackage[utf8x]{inputenc}
\usepackage[english]{babel}
\usepackage{times}
\usepackage{array}
\usepackage{amsmath}
\usepackage{graphicx}
\usepackage{amssymb}	
\usepackage{amsfonts}
\newcommand{\lambdabar}{{\mkern0.75mu\mathchar '26\mkern -9.75mu\lambda}}

\newcommand{\hl}{\lambdabar}

\newcounter{ourcount}
\title[Fast radio bursts]{On the nature of fast radio bursts}
\author[Ya.~N. Istomin]{Ya.~N. Istomin $^{1,2,3}$ \\
${1}$ P.N.~Lebedev Physical Institute, Leninsky Prospect 53, Moscow 119991, Russia \\
${2}$ Moscow Institute Physics and Technology, Institutskii per. 9, Dolgoprudnyi, Moscow region, 141700, Russia \\
${3}$ E-mail: istomin@lpi.ru \\}

\begin{document}
\date{}
\pagerange{\pageref{firstpage}--\pageref{lastpage}}
\pubyear{2017}
\maketitle
\label{firstpage}

\begin{abstract}
Scenario of formation of fast radio bursts (FRBs) is proposed. Just like radio pulsars, sources of FRBs are magnetized
neutron stars. Appearance of strong electric field in a magnetosphere of a neutron star is associated with close passage of a dense
body near hot neutron star. For the repeating source FRB 121102, which has been observed in four series of bursts, the period of orbiting of the body is about 200 days. Thermal radiation from the surface of the star (temperature is of the order of $ 10^8 \, K $) causes evaporation and ionization of the
matter of the dense body. Ionized gas (plasma) flows around the magnetosphere of the neutron star with the velocity $ u \simeq 10^7 \, cm / s $, and creates electric potential $ \psi_0 \simeq 10^{11} \, V $ in the polar region of the magnetosphere. Electrons from the plasma flow are accelerated toward the star, and gain Lorentz factor of $ \simeq 10 ^ 5 $. Thermal photons moving toward precipitating electrons are scattered by them, and produce gamma photons with
energies of $ \simeq 10^5 \, m_e c^2 $. These gamma quanta produce electron-positron pairs in collisions with thermal photons. The multiplicity,
the number of born pairs per one primary electron, is about $ 10^5 $. The electron-positron plasma, produced in the polar region of magnetosphere,
accumulates in a narrow layer at a bottom of a potential well formed on one side by a blocking potential $ \psi_0 $, and on the other side by pressure of thermal radiation. The density of electron-positron plasma in the layer increases with time, and after short time the layer becomes a mirror
for thermal radiation of the star. The thermal radiation in the polar region under the layer is accumulated during time $ \simeq 500 \, s $, then the plasma layer is ejected outside. The ejection is observed as burst of radio emission formed by the flow of relativistic electron-positron plasma.

\end{abstract}

\begin{keywords}
radiation mechanisms: general -- stars: neutron
\end{keywords}

\section {Introduction}
                                       
Fast radio bursts (FRBs), first discovered in 2007 by Lorimer et al., are short radio signals of several
milliseconds duration ($ \tau \simeq 10^{-3} \, s $) with the energy flux 
from tens of milliJansky to several Jansky in the radio band of $ \simeq 1-10 \, GHz $. 
The main feature of FRB is
high observed dispersion measure DM. It is several times greater than the maximum 
value of the dispersion measure
of radio signal passing through our Galaxy. High DM suggests that the radio signal 
propagates through intergalactic medium.
Also the rotation measure RM, connected with presence of a magnetic field in a plasma,
through which a radio signal propagates, is high.
The ratio of the rotation measure to the dispersion measure, $ RM / DM $, equals to the 
average value of the longitudinal
magnetic field on the line of sight. On the contrary, this ratio is one order of magnitude smaller 
than the average magnetic field in the 
Galaxy, which also indicates that the radio signal propagates in
intergalactic medium. Finally, the repeated source of FRB 121102 was identified
with a galaxy having a redshift of $z = 0.192 $ (Chatterjee, et al., 2017).
Assuming that large values of DM and RM are achieved in the
intergalactic medium implies
cosmological distances to sources of FRB, $ z \simeq 1 $. If we assume isotropic radiation into the solid angle
$\simeq 4\pi$ and cosmological distance to sources, 
the radiated power in the radio
band should reach value of the order of $ 10^{43} \, erg/s $ and the total radiated energy 
of $ 10^{40} \, erg $. The beaming of FRB can decrease these estimates by many orders of magnitudes (Katz, 2017a). If we do not take into account the relativistic
time delay in the source, then the source size $ l $ is estimated to be of the order of 
$ 3 \cdot 10^7 \, cm $. Further the energy density in the source 
is equal to $ 3 \cdot 10^{17} \, erg/cm^3 $ in the radio range only. This corresponds to the 
electric field $ E \simeq 10^{12} \, V / cm $, which is at least
two orders of magnitude higher than the value of the atomic field, $ E_a \simeq 0.5 \cdot 10^{10} \, 
V/cm $. At such large energy densities,
we have to expect radiations in optical, x-ray and gamma ranges, which are not observed.
It is difficult to understand what the extragalactic source could be. So the nature of FRB remains unknown today.

There are several different consequences regarding the origin of FRB. First of all, because of 
smallness of $l$ the source of FRB is probably a neutron star. 
It is known that
neutron stars are observed as radio pulsars, x-ray pulsars, magnetars, rotating radio 
transients (RRATs). High
brightness temperature of FRB, up to $10^{37} \, K$, suggests a coherent mechanism of radio 
emission similar to the one of radio pulsars,
especially the mechanism of radiation of giant radio pulses, observed from several radio pulsars (Soglasnov et al., 2004; Hankins and Eilek, 2007). When radio 
sources of this kind with the same radiation mechanism are placed at cosmological
distances it is necessary to assume extreme values of neutron star 
parameters: large magnetic fields on its
surface, $ B_0 \simeq 10^{15} \, G $, and
fast rotation, $ \Omega_{NS} \simeq 10^4 \, s^{-1} $. 
But neutron stars with so extreme parameters lose their rotation energy quite rapidly, during a time of the order of $ 10^3 \, s $,
and can not give repeated bursts. As for magnetars, in particular SGRs, the energy is stored in a magnetic field inside the star,
and they produce bursts in gamma and x-ray energy ranges, which are not observed in FRBs. The energy released in a burst can be
the energy of the electric field in the rotating magnetosphere of a neutron star (Katz, 2017b). The author calls such phenomenon 'pulsar lightning'. However,
it is not clear how and how fast transition from one configuration of the magnetosphere to another is possible. Earlier, the term 'lightning' was used for
explanation of radiation of RRATs (Istomin and Sobyanin, 2011a,b). It was specifically shown that cascade formation of electron-positron pairs and their acceleration by a longitudinal electric field in the polar vacuum magnetosphere, caused by absorption of an energetic gamma quantum from Galactic and extragalactic backgrounds, should look like a flash of lightning on Earth during a thunderstorm.

The closest to what is discussed in this article are models of FRBs origin by the interaction of neutron stars with other bodies (planets,
comets, asteroids). These are primarily direct collisions of bodies with a neutron star (Di and Dai, 2017; Dai et al., 2016) and the interaction of a relativistic pulsar wind with a companion of a neutron star (Mottez and Zarka, 2014).

However, apparently, FRB is a different class 
of sources of radio emission, which are neutron stars. This can be seen from analysis of observations of the source FRB 121102.

\section {FRB 121102}

The large energy density in a source most likely suggests cataclysmic event, such as an explosion, 
and hence destruction of the
source. Indeed, no recurrent events were recorded until recently. One radio source
FRB 121102, discovered in November 2012, flared up ten times within
16 days ($ 1.4 \cdot 10^6 \, s $) 926 days later (May 2015).
With the exception of one long time interval between consecutive flashes, also by the way about 16 days, 
the time interval between bursts were random from
$ \simeq 20 $ seconds to $ \simeq 1 $ hour. The average duty cycle was 
$ \simeq 670 \, s \simeq 11 \, min $. Then after 164 days
(November 2015) six more bursts were recorded during 25 days ($ 2 \cdot 10^6 \, s $) with two breaks 
of 6 and 18 days and with the average
duty cycle of $ 430 \, s $. In September 2016 (after 287 days) the source FRB 121102 flashed four 
more times with the average time between bursts
$ \simeq 10^3 \, s $. Finally, after 340 days fifteen new bursts were recorded with average
duty cycle of $ 150 \, s $ in August 2017.

 We see that in the temporal activity the FRB 121102, apart from the duration 
of the radio emission pulses, $ \tau \simeq 10^{-3}
\, s $, exhibits three characteristic times: 1) $ \tau_1 \simeq 500 \, s $ is
the average time interval between consecutive bursts; 
2) $ \tau_2 \simeq 20 \, d \simeq 2 \cdot 10^6 \, s $ is the duration of
series of bursts and the duration of
continuous breaks between bursts in the series; 3) $ \tau_3 \simeq 200 \, d \simeq 2 \cdot 10^7 \, s $ 
is the average
time between series of bursts. These times are very different, $ \tau_1 << \tau_2 << \tau_3 $.

Thus, the source FRB 121102 exhibits activity during the time $ \tau_2 = 2 \cdot 10^6 \, s $ 
in the form of short bursts of
few milliseconds in duration. The values of duration of radio emission, intensity and duty cycle are close to 
the same values observed from
so-called rotating radio transients (RRATs). However, although the duty cycle of RRATs isa random 
value, it is a multiplier of 
some constant time unit. This time unit is period of rotation of a neutron star, which increases
with time like that for 
radio pulsars. The value of time derivative of the period, $ \dot{P} $, allows us to estimate the 
magnetic field strength on the stellar surface.
It is $ B \simeq 10^{12}-10^{13} \, G $, which is similar to radio pulsars. RRATs rotate slower than 
radio pulsars (the period of rotation is of the order 
of $ \simeq 10 \, s $). Because of this there is no permanent generation of electron-
positron plasma in the neutron star magnetosphere. Although the rotation of magnetic field, frozen into the star, generates the 
electric field in the magnetosphere, $ E \simeq
v_{rot} B/c $, the electric field is not sufficient for the continuous production of electron-positron plasma. 
However, strong magnetic field and
the electric field, induced by the rotation of magnetosphere of the neutron star, can lead to the burst of 
the production of electron-positron plasma under some external action.
Such external action in the case of RRATs is the Galactic and extragalactic gamma rays with energies 
above $ 1 \, MeV $ according to mechanism of RRATs  
developed by Istomin and Sobyanin (2011a,b). Now a natural question arises: what will 
happen if magnetized neutron star rotates 
even more slowly ($ P> \tau_1 \simeq 500 \, s $)
than the neutron star which is the source of RRAT? There is strong magnetic 
field in its magnetosphere, but there is practically absent
the electric field, which is necessary for acceleration of electrons and positrons in the 
magnetosphere and for beginning of a cascade plasma production process.

The electric field, $ E \simeq uB/c $, can also arise in the magnetosphere when a sufficiently 
dense flow of charged
particles moves through the magnetosphere with the velocity $ u $. Thus, we come to the conclusion that if the source of
FRB is a magnetized neutron star, which rotates slowly enough, and
the birth of a relativistic plasma occurs in its magnetosphere, an external action is 
necessary to distort the magnetosphere and to create
the electric field. The presence of a neutron star as a source of FRB is indicated by short duration of the burst of radio 
emission.  
Such an effect can be a flow of a sufficiently dense plasma. Note that 
from one flash to another flash the  dispersion measure
is not constant but varies within $ 3 \% $ during the time
$ \simeq \tau_1 = 500 \, s $. This would correspond to the motion of electron density inhomogeneities 
of the size of $ <10^{13} \ cm $. This size is too small either
for the Galaxy, or for mentioned intergalactic medium, but reflects the presence of plasma in the 
immediate vicinity of the neutron star.
The characteristic values of plasma flow scale $ L $ and its velocity $ u $ can be estimated from 
the following relations. First, it is $ L / u = \tau_2 =
2 \cdot 10^6 \, s $. Second, we assume that the repetition time of the burst 
series $\tau_3$ is the orbiting period a dense body
(planet, comet, asteroid) around the neutron star, $ P_{orb} = \tau_3 $. Since $ \tau_2 << \tau_3 $, 
the orbit of a body is strongly elongated. Therefore
$ u^2 L = 2GM_{NS} = 3.7 \cdot 10^{26} \, cm^3/s^2 $. Here $G$ is the gravitational constant, 
and $ M_{NS} $ is the mass of the neutron star, which
we put equal to $ 1.4 \, M_{\odot} $. Thus, we have
\begin{equation}
L=10^{13}\, cm, \, u=6\cdot 10^6\, cm/s.
\end{equation}      

\section{Electric field}

As a result, the scenario of interaction of a dense body orbiting around the magnetized 
hot neutron star (for FRB 121102),
looks like that: close pass of the body at the distance less than $ 1 \, a.u. $ causes 
evaporation from the body, and dense plasma
flow around the magnetosphere of neutron star perturbing magnetic field in the polar region, 
and generating the longitudinal electric field. For the temperature
of the surface of the neutron star $ T_{NS} $ of the order of $ 10^8 \, K \simeq 10 \, KeV $, 
the energy flux of x-ray photons is
$ S_{NS} = 7 \cdot 10^{40} \, erg/s $ (at the stellar radius of
$ R \simeq 10 \, km $), which exceeds the solar luminosity by $ 2 \cdot 10^7 $ times. The evaporated 
plasma has the temperature on the order of the temperature of
evaporation of a solid body $ T_b \simeq 0.1 \, eV $, and its
thermal velocity $ v_T $ is equal to $ (2T_b/Am_p)^{1/2} \simeq 10^5 \, cm/s $. Here we chose the mean 
atomic number of the evaporated ions to be of the
order of $ A = 20 $, and $ m_p $ is the proton mass. Thus, $ v_T << u $, and the velocity 
of the flow around the neutron star magnetosphere is approximately
equal to $ u $.
It should be noted that the temperature of the neutron star should not be too large, such that the 
radiation force acting on the ionized gas does not exceed
the gravitation force, $ T_{NS}<(Am_p cGM_{NS} / ZR^2 \sigma \sigma_T)^{1/4} = 0.3 \cdot 10^8 (A/Z)^{1/4} R_6^{- 1/2} \, K $. 
Here $ \sigma $ is
the Stefan-Boltzmann constant, and $ \sigma_T $ is the Thomson scattering cross-section.
The pressure of the plasma flow destroys the magnetic field of the neutron star at distances 
from the center of the star larger than a certain distance $ r $
determined by equality of pressures, $ Am_pn_eu^2/2 Z = B(r)^2/8\pi $. The value of $ Ze $ 
is the average ion charge, $ A / Z \simeq 10-20 $.
We obtain $ B (r) =
(8 \pi Am_p n_e u^2/Z)^{1/2} $. The region of perturbed magnetosphere at the distance $ r $ from 
the star, called casp, has size of the same $ r $.
However, this size decreases to the polar oval of small size on the stellar surface. 
Electron-positron plasma fills this polar region by a cascade process described below.
The magnetic field at the level $ r $
can be found from that the magnetic field of the neutron star at large distances 
is dipole, $B=B_0(r/R)^{-3}$.
Here $B_0$ is the value of the magnetic field on the surface of NS, 
$ B_0 = 10^{12} B_{12} \, G, \, B_{12} = B_0/10^{12} G $. Thus,
$B(r) = 10^{12} B_{12} (r/R)^{-3} \, G \simeq 10^{3} \, G $ for $ r = 10^3 R $. The electric 
field induced in the polar region, $ E = uB/c $,
is equal to $ 6 \cdot 10^{10} B_{12} (r/R)^{-3} \, V/cm \simeq 60 \, V/cm $ for $ r = 10^3 R $. 
Accordingly, the arising
voltage in the cusp is equal to $ |\psi_0| = Er = 6 \cdot 10^{16} B_{12} R_6 (r/R)^{-2} \, V \simeq 6 \cdot 10^{10} \, V $ for $ r = 10^3 R $
($ R_6 = R/10^{6} \, cm $). The electric field originates in the open region of the magnetosphere 
of the neutron star, in its polar region
around the axis of a stellar magnetic moment $ {\bf M} $. On the surface of the star this region 
is almost a circle of the radius 
$ R_0 \simeq R(R/r)^{1/2} \simeq 3 \cdot 10^4 \, cm $. This polar circle is
the same as in neutron stars, which are radio pulsars. Knowing the magnitude of the magnetic 
field $ B(r) $, we can determine the electron density
$ n_e $ in the incoming plasma flow, $ n_e = n_0 = B^2(r) Z/8 \pi Am_pu^2 \simeq 10^{13} \, cm^{-3} $ 
for $ r = 10^3 R $.
A small fraction of electrons from the incoming stream penetrates into the polar region of the 
magnetosphere. Their flux $ S $ is equal to 
$ S = \pi \eta n_0 u r^2 \simeq 2 \cdot
10^{39} \eta \, s^{-1} $. Here $ \eta $ is the efficiency of penetration of electrons into the 
polar magnetosphere, $ \eta << 1 $.

The electric field arising in the magnetosphere at the distance $ \simeq r $ penetrates deep into 
the magnetosphere in the polar region. Its dependence on
of the height $ h $ above the surface of the star can be determined by solving the Laplace equation 
in the region bounded by the surface of the radius $ \rho(r) = h(h/r)^{1/2} $
with boundary conditions: $ \psi(h = r) = \psi_0, \psi(h = R) = 0, \psi (h, \rho = h (h/r)^{1/2}) = 0, $
$$
\frac{\partial^2\psi}{\partial h^2}+\frac{1}{\rho}\frac{\partial}{\partial\rho}\left(\rho\frac{\partial\psi}{\partial\rho}\right)=0.
$$  
Replacing the transverse derivative $ \Delta_{\rho} \psi $ by $ - (4r/h^3) \psi $, we arrive at the 
equation
$$
\frac{\partial^2\psi}{\partial h^2}=\frac{4r}{h^3}\psi.
$$
The solution of this equation is $ \psi = const \cdot (h/r)^{1/2} K_1[4 (h/r)^{-1/2}] $, where $ K_1(x) $ 
is the McDonald function of the first order. Since $ h/r <1, \, x>4 $,
one can use asymptotic presentation of the McDonald function for large arguments. As a result, 
we have
\begin{equation}\label{psi} 
\psi(h)=\psi_0\left(\frac{h}{r}\right)^{3/4}\exp\left[-4\left(\frac{r}{h}\right)^{1/2}+4\right].
\end{equation}
For the practical purpose the asymptotic presentation (\ref{psi}) does not differ from the exact solution 
expressed by the McDonald function,
and we will use the expression (\ref{psi}) below.
We see that the longitudinal electric field exists only in the upper part of the polar 
tube, $ r> h> r^* \simeq 0.6 r $. The field is
exponentially suppressed when approaching the star,. Thus, electrons falling into the polar region at $ h \simeq r $ are accelerated 
toward the star, $ \psi_0<0 $,
and get relativistic energy equal to $ e|\psi_0| $, i.e. their Lorentz factor becomes equal to 
$ \gamma_0 = 1.2 \cdot 10^{11}B_{12} R_6(r/R)^{-2}
\simeq 10^5 $.

\section {Production of electron-positron plasma} 

Thermal photons with energies $ E_{ph} = m_ec^2 \epsilon $ propagating from the stellar surface 
are scattered by relativistic electrons
and produce high-energy photons with energies $E_{ph}^\prime = m_ec^2 \epsilon^{\prime}$,
\begin{equation}\label{epsilon}
\epsilon^{\prime}=\epsilon\frac{4\gamma^2}{1+\gamma^2\theta^2+4\gamma\epsilon}.
\end{equation}
Here the angle $ \theta $ is the angle of propagation of a scattered photon with respect to the 
velocity of the relativistic electron. 
Since, as can be seen from
(\ref{epsilon}), $ \epsilon^{\prime} $ is large for photons scattered in the direction of 
propagation of the electron, 
$ \theta \simeq \gamma^{-1} << 1 $, then
we use the approximation $ 1-\cos\theta \simeq \theta^2/2 $. If we neglect the electron recoil, 
$ 4 \gamma \epsilon <1 $, then the maximum energy
of the scattered photon ($ \theta = 0 $) is $ \epsilon^{\prime}_m = 4\gamma^2 \epsilon $. 
For the Planck spectrum with the temperature of $ T_{NS} $ the maximum
energy density is at the photon energy $ E_{ph} = 2.82 T_{NS} = 24T_8 \, KeV $. Here 
$ T_8 $ is the stellar surface temperature
in units of $ 10^8 \, K $, $ T_8 = T_{NS}/10^8 K $. Thus, the characteristic value of the 
thermal photon energy is
$ \epsilon \simeq 4.7 \cdot 10^{- 2} T_8 $. The value of $ 4 \epsilon \gamma $ is equal to 
$ 2 \cdot 10^4 T_8 $. This means that for small scattering angles,
$ \theta <\theta^* =
2(\epsilon/\gamma)^{1/2} \simeq 1.4 \cdot 10^{-3} T_8^{1/2} $, the electron completely loses its 
energy, $ \epsilon^\prime = \gamma $.
At larger angles, $ \theta> \theta^* $, the energy of the scattered photon is $ \epsilon^\prime = 4\epsilon/\theta^2 <\gamma $. 
The scattered photon $ \epsilon^\prime $
propagates toward the star where the density of thermal photons increases $\propto h^{-2} $, and the magnetic field also grows $\propto h^{-3}$. 
Therefore it is possible
to create electron-positron pairs by two ways: 1) in collisions of scattered photons with thermal photons moving towards, 
$ \epsilon^\prime + \epsilon = \gamma^- + \gamma^+ $, and 2) birth in magnetic field, when the scattered photon intersects the line of 
strong magnetic field at the angle $\beta$,
$ \epsilon^\prime \sin \beta> 2 $. It should be noted that in contrast to radio pulsars, energetic electrons $ \gamma $ and photons 
$ \epsilon^\prime $ propagate toward
the star, where the birth of pairs is more efficient, but not outwards. In addition, curvature photons, which play a major role in 
initiation of the cascade
production of pairs in magnetospheres of radio pulsars, have small energies here and are incapable to produce pairs. Indeed, the energy 
of a curvature photon $\epsilon_c$ is
$ \epsilon_c = \hl\gamma^3 / \rho_c \simeq 5 \cdot 10^{-4} << 1 $. Here $ \hl $ is the Compton wavelength of electron,
$ \hl = \hbar/m_ec = 3.86 \cdot 10^{-11} \, cm $, $ \rho_c $ is the radius of curvature of magnetic field lines in the polar region,
$ \rho_c = 4(rh)^{1/2}/3> 4 (rR)^{1/2}/3 \simeq 7 \cdot 10^7 \, cm $. 

Let us consider the process of production of electron-positron pairs by scattered photons:

1) For the production of a pair in collision of a scattered photon with a thermal photon, $ \epsilon^\prime \epsilon> 1 $, taking 
into account (\ref{epsilon}), we obtain condition
$$
\theta <[(2\epsilon-1/\gamma)^2-2/\gamma^2]^{1/2} \simeq 2\epsilon, \, \epsilon>\epsilon_1=(2^{1/2}+1)/2\gamma.
$$
The inverse Compton scattering cross-section in the laboratory coordinate system, associated with the neutron star, has the form (Berestetskii, Lifshitz and Pitaevskii, 1982)
\begin{eqnarray}\label{sigma}
d\sigma=\frac{8\pi r_e^2\gamma^2\theta d\theta}{(1+\gamma^2\theta^2+4\epsilon\gamma)^2}\left[\frac{1}
{(1+\gamma^2\theta^2)^2}-\frac{1}{1+\gamma^2\theta^2}+ \right. \\ \nonumber
\left.\frac{1}{4}\left(\frac{1+\gamma^2\theta^2
+4\gamma\epsilon}{1+\gamma^2\theta^2}+\frac{1+\gamma^2\theta^2}{1+\gamma^2\theta^2+4\gamma\epsilon}\right)\right].
\end{eqnarray}
Here $ r_e $ is the classical radius of an electron, $ r_e = 2.82 \cdot 10^{-13} \, cm $. Integrating the cross-section (\ref{sigma}) over 
the angle $ \theta $ from $ 0 $ to
$ [(2\gamma \epsilon-1)^2-2]^{1/2} / \gamma $, and then averaging over the Planck spectrum,
we obtain the value of the thickness $ \tau $ gained by an electron moving toward the neutron star when it is scattered by thermal photons
\begin{equation}\label{tau}
\tau=\int_h^{r^*}dh'\int_{\epsilon_1}^\infty\frac{dn_{ph}}{d\epsilon}d\epsilon\int_0^{2\epsilon}\frac{d\sigma}{d\theta}d\theta=
\frac{4a}{\pi}\frac{r_e^2 T^2R}{\hl^3\gamma}\left(\frac{R}{h}-\frac{R}{r^*}\right).
\end{equation}
The coefficient $ a $ is proportional to the integral of the scattering cross-section (\ref{sigma}) over the Planck spectrum,
starting from the photon energies $ \epsilon> (2^{1/2} +1) / 2 \gamma $ to infinity. It depends on the electron energy $ \gamma $ and the 
temperature of the star $ T $, $T=T_{NS}/m_e c^2$.
For parameters of interest, $ T_8 \simeq 1, \, \gamma \simeq 10^5 $, the value of $a$ is $ a \simeq 1 $. The coefficient 
$ 4ar_e^2 T^2R / \pi\hl^3 \gamma $ in the right hand side of
(\ref{tau}) is large, $ \simeq 5 \cdot 10^3 T_8^2 \gamma_5^{-1} R_6 \, (\gamma_5 = \gamma_0 / 10^5, \, R_6 = R / 10^6 \, cm) $. It means
that the thickness $ \tau $ becomes of the order of unity fairly fast, $(r^*-h)/r^*\simeq 0.12$,
and the electron completely loses energy, emitting a 
gamma quantum with the energy $ \epsilon^\prime \simeq \gamma $.

Let us now find the efficiency of production of electron-positron pairs produced by collisions of energetic photons $ \epsilon^{\prime} $ with 
thermal photons $ \epsilon $.
The cross-section for pair production is (Berestetskii, Lifshitz and Pitaevskii, 1982)
$$
\sigma^{\pm}=\frac{\pi r_e^2}{2}(1-v^2)\left[(3-v^4)\ln\frac{1+v}{1-v}-2v(2-v^2)\right],
$$
where $ v $ is equal to $ v = (1-1/ \epsilon\epsilon^{\prime})^{1/2} $. Again averaging over the Planck spectrum,
starting from the energy $ \epsilon = 1 / \epsilon^\prime $, and assuming $ \epsilon^{\prime} \simeq \gamma $, we get the value of the 
thickness $ \tau_e $ gained by the gamma quantum of the
energy of $ \epsilon^\prime $ with respect to the production of a pair,
\begin{equation}
\tau_e=\frac{b}{2\pi}\frac{r_e^2T^2R}{\hl^3\gamma}\left(\frac{R}{h}-\frac{R}{h_i}\right).
\end{equation}
The numerical coefficient $ b $ for $ T_8 = 1, \, \gamma = 10^5 $ is $ b \simeq 25 $. Here $ h_i $ is the initial height from which the 
gamma photon begins to
move toward the star. We see that the birth of an electron-positron pair is even more effective, $ b / 8a \simeq 3 $ times, than gamma-ray 
radiation by an
electron. Thus, the primary electron, passing the distance $ \simeq qr^*, \, q \simeq 0.12 (1 + 1/3) = 0.16 $ through the thermal radiation, 
produces a pair 
with electron and positron energies $ \simeq \gamma/2 $. In turn, a born pair produces a new pair in the photon field of the star, etc. So
the cascade production of electrons and positrons occurs. The minimum number of possible cascades $K$, if we do not take into account decrease of the energy
of secondary electrons ($ \gamma $ in the expression (\ref{tau})), is determined by the condition
$$
q\sum_{n=1}^K(1-q)^n=1-\frac{R}{r^*}.
$$
Summing up, we obtain $K + 1 = [\ln (R / r^*) / \ln(1-q)] $. Substituting characteristic values $ r^* \simeq 6 \cdot 10^2R, \, q \simeq 0.16 $, 
we find $K \simeq 35 $.
Since $ 2^{35} \simeq 3 \cdot 10^{10} >> \gamma_0 $, the number of pairs $ \lambda $ generated by one primary electron is determined by the relation
$ \lambda \simeq \gamma_0 / 2 $,

2) Now we consider the single-photon production of electron-positron pairs by scattered photons $ \epsilon^\prime $ in the magnetic field of the neutron star.
We need to know the number of generated photons with energy $ \epsilon^\prime> 2 $. They are larger than the previously calculated density of
energetic photons $ \epsilon^\prime> \epsilon^{-1} \simeq 10^2 $, which collide with thermal photons to produce pairs. Expression (\ref{epsilon}) for
the energy of scattered photons determines regions of angles $ \theta $ and energies $ \epsilon $ of primary photons,
$$
\theta<(2\epsilon-4\epsilon\gamma^{-1}-\gamma^{-2})^{1/2}\simeq(2\epsilon)^{1/2}, \, \epsilon>1/2\gamma(\gamma-2)\simeq 1/2\gamma^2.
$$
As before, using the cross-section (\ref{sigma}), integrating over the Planck spectrum, we obtain expression for the thickness $\tau_1$,
\begin{equation}
\tau_1=\frac{4a_1}{\pi}\frac{r_e^2 T^2R}{\hl^3\gamma}\left(\frac{R}{h}-\frac{R}{r^*}\right),
\end{equation}
where the constant $ a_1=3.8 $ in the region of parameters $ T_8 \simeq 1, \, \gamma \simeq 10^{5}$. Thus, the mean free path $ l_e $ of fast
electron with respect to the production of secondary photons with energy $ \epsilon^\prime> 2 $ is 3.8 times smaller than the mean 
free path for production of energetic 
photons with energies $ \epsilon^\prime> 10^2, \, l_e \simeq 3 \cdot 10^{-2} r^* $.

Scattered secondary photons emitted by electrons along magnetic field lines, and propagating toward the surface of the star,
begin to cross magnetic field lines at an angle $ \beta \ne 0 $ due to curvature of magnetic field lines in the polar region of the magnetosphere.
Thus, the angle $ \beta $ is equal to
$$
\beta=\int_h^{h_i}\rho^{-1}_c(h')dh'=\frac{3}{2}\left[\left(\frac{h_i}{r}\right)^{1/2}-\left(\frac{h}{r}\right)^{1/2}\right].
$$
Here the value of $ h_i $ is equal to the initial altitude, from which the photon begins to propagate. As we will see pair production occurs below the height $ h \simeq 10R, \, h << r $. Because of this and because of strong dependence
of the magnetic field strength on the height, $ B \propto h^{-3} $ angle
$ \beta $ can be considered to be a constant, $ \beta \simeq 3h_i^{1/2} / 2r^{1/2} $. The probability of pair production in the magnetic 
field per unit time is (Berestetskii, Lifshitz and Pitaevskii, 1982)
\begin{equation}
w=\frac{3^{3/2}\alpha c}{2^{9/2}\hl}b|\sin\beta|\exp\left\{-\frac{8}{3\epsilon^\prime b|\sin\beta|}\right\}\Theta(\epsilon^\prime|\sin\beta|-2).
\end{equation}
The quantity $ b $ is the magnetic field intensity in units of the critical field, $ b = B / B_\hbar, \, B_\hbar = 4.4 \cdot 10^{13} \, G $, $ \alpha $ is 
the fine-structure constant, $ \alpha = 1/137 $, $\Theta(x)$ is the stepwise theta function. The height $ h $, at which a pair is born with the 
probability $ \simeq 1 $, is determined by the condition
$$
\frac{1}{c}\int_h^{h_i}w(\epsilon^\prime, h')dh'=1.
$$
Setting $ \sin \beta \simeq \beta $, we obtain
\begin{equation}
\left(\frac{h}{R}\right)^3=\frac{9}{16}\epsilon^\prime b_0\left(\frac{h_i}{r}\right)^{1/2}\Lambda,
\end{equation}
where the value of $b_0$ is $b_0=B_0/B_\hbar=2.3\cdot 10^{-2}B_{12}$, and 
\begin{eqnarray*}
&&\Lambda=\ln\left\{\frac{3^{1/6}}{2^{17/6}}\frac{R\alpha}{\hl}(\epsilon^\prime)^{-2/3}b_0^{1/3}\left(\frac{h_i}{r}\right)^{1/6}\right\}  \\ 
&&-\frac{5}{3}\ln\ln\left\{\frac{3^{1/6}}{2^{17/6}}\frac{R\alpha}{\hl}(\epsilon^\prime)^{-2/3}b_0^{1/3}\left(\frac{h_i}{r}\right)^{1/6}\right\}.
\end{eqnarray*}
The characteristic values of $ \Lambda$ are $\simeq 15-20$. We see that photons with energy $ \epsilon^\prime \simeq 5 $ produce electron-positron pairs only
near the stellar surface, $ h \simeq R $, while energetic photons, $ \epsilon^\prime \simeq 10^5 $, can produce pairs
at a distance $ h / R \simeq  30 $.
After pairs are born, they lose their transverse momenta, emitting synchrotron photons with energy
$ \epsilon_s \simeq 3b \gamma^2 |\sin \beta| / 2 = 3b_0 / 2 |\sin\beta| = 3b_0 (r / h_i)^{1/2} /2
\simeq 1 $. They can not produce new pairs. Thus, the cascade single-photon production of electron-positron
pairs in a strong magnetic field
in our case is actually absent, in contrast to the cascade production of pairs in magnetospheres of radio pulsars.

\section{Plasma trap}

Thus, we see that primary electrons with the Lorentz factor of $ \gamma_0 \simeq 10^5 $ effectively produce electron-positron pairs by directly interacting
with thermal photons. In view of the cascade character of process, most of them
have multiplicity $ \lambda \simeq 10^5/2 $ and energies of $ \gamma \simeq 1 $ with the energy spread of $ \Delta \gamma \simeq 1 $.
However, they do not reach the surface because the flux of thermal photons from the surface pushes them out. As a result,
electrons and positrons are accelerated outward from the star,
$$
\frac{d\gamma}{dt}=\sigma_T\frac{\pi^2cT^4}{60\hl^3}\left(\frac{R}{h}\right)^2.
$$
Here $ \sigma_T $ is the Thomson cross-section, $ \sigma_T = 8 \pi r_e^2/3 $, and we substitute $ \sigma=\pi^2/60 \hbar^3 c^2 $ for the 
Stefan-Boltzmann constant in energy units. Integrating and setting $ d\gamma /dt = cd \gamma / dh $, we obtain
$$
\gamma=\frac{2\pi^3}{45}\frac{r_e^2 T^4R}{\hl^3}\left(1-\frac{R}{h}\right).
$$
In the region $ R<h<r^* $, where there is factually no electric field (\ref{psi}), secondary electrons and positrons get the energy $ \gamma_f $,
$$
\gamma_f=\frac{2\pi^3}{45}\frac{r_e^2 T^4R}{\hl^3}=1.6\cdot 10^5 T_8^4R_6.
$$
Positrons freely escape outside freely, while electrons obeying the condition
$$
\kappa = 0.6 \gamma_5 / T_8^4 R_6> 1
$$
are reflected from the electric potential (\ref{psi}) and become trapped. This trap is formed from one side by the electric
potential created by the plasma flow in the 
magnetosphere, and the thermal radiation of the star itself from the other side. Electrons stop at the distance
$ h = h_r $ from the star defined by the condition $ \gamma_f = e|\psi (h_r)|/m_e c^2 $,
$$
h_r=r\left[1+\frac{1}{4}\ln\kappa+\frac{3}{8}\ln\left(1+\frac{1}{4}\ln\kappa\right)\right]^{-2}.
$$
After reflection electrons begin to move toward the surface of the star under the action of the trapping potential $ \psi(h) $, and, if
it were no interaction with thermal photons moving toward them, they would come to the stellar surface with the energy corresponding to
the Lorentz factor of $ \gamma_0 / \kappa $. However, they emit gamma quanta, which produce new electron-positron pairs, are decelerated
to nonrelativistic energies and are  picked up again by the radiation of the star. The force acting on electrons moving outward from the star is potential
since the cross-section for scattering of thermal photons is equal to the Thomson cross-section, which does not depend on the electron energy. Corresponding
potential $ \psi_T $ is equal to
$$
\psi_T = m_e c^2\gamma_f\frac{R}{h}.
$$
The plot of the potential $ \psi_T / e |\psi_0| \, (\kappa=3)$ is shown in Figure 1. Figure also shows the potential of the electric field $ \psi (h) / \psi_0 $ (Eq. \ref{psi}) and the total potential $ (e |\psi(h)| + \psi_T) / e |\psi_0| $.
\begin{figure}
\begin{center}
\includegraphics[width=6cm]{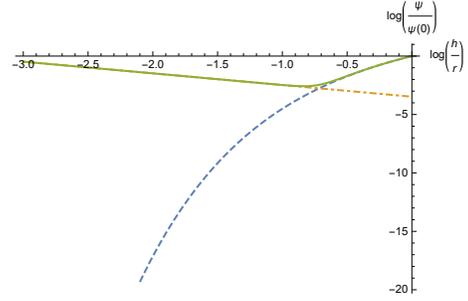}
\end{center}
\caption{The logarithms of the potential $\psi_T$ acting on electrons from the thermal stellar radiation (dot-dash line), of the potential $e|\psi(h)|$ (Eq. \ref{psi}) (dotted line) and of
the total potential $ e |\psi(h)| + \psi_T $ (solid line) as functions of the height $h$.}
\end{figure}
As a result of pair production, electrons are accumulated at the bottom of the potential well formed by the total potential $ e |\psi (h)| + \psi_T (h) $.
The location of the bottom, $ h_m $, is equal to (with logarithmic accuracy)
\begin{eqnarray}\label{hm}
&&h_m=r\left[1+\frac{1}{4}\ln\kappa+\frac{1}{4}\ln\left(\frac{2r}{R}\right)\right. \\  \nonumber
&&\left. -\frac{5}{8}\ln\left(1+
\frac{1}{4}\ln\kappa+\frac{1}{4}\ln\left(\frac{2r}{R}\right)\right)\right]^{-2}<h_r,
\end{eqnarray}
Thus, a primary electron with energy $ \gamma_0 $, after almost reaching the surface of the star, and losing its energy, returns 
back, receiving from thermal photons energy less than the original one, $ \gamma_0 / \kappa $, then again moves toward the star and again produces quanta and
pairs until it settles near the height of $ h_m $. The multiplicity, i. e. the number of generated electrons per one primary electron, increases as $ \lambda
\simeq \gamma_0 \kappa / 2 (\kappa-1) $. It should be noted that the decelerating force acting on an electron when it moves toward the star is not conservative,
since the scattering cross-section is inversely proportional to the electron energy, $ \propto \gamma^{-1} $, as can be seen from  expression (\ref{tau}). 
Therefore, the motion of energetic electrons toward the star is not the motion
in the total potential $ e |\psi (h)| + \psi_T (h) $. Their deceleration occurs faster, $ d\gamma^2/dh \propto h^{-2} $, but 
not $ d \gamma / dh \propto h^{-2} $
as for the motion in the potential $ \psi_T $. But for subrelativistic electrons, $ \Delta \gamma \simeq 1 $, located near the bottom
of the potential well, the cross-section for scattering of thermal photons is the Thomson one, and their motion is potential in the 
potential $ e |\psi| + \psi_T $.

The thickness of the layer $ \Delta h $, where electrons are accumulated, is small. It
is determined by the spread of electron energies near the surface of the star after production of gamma-quanta and pairs, $ \Delta \gamma \simeq 1 $,
$$
\Delta h=h_m\frac{\Delta\gamma}{2\gamma_f}\left(\frac{h_m}{r}\right)^{1/2}<<h_m.
$$
This layer creates an additional electric potential $\psi_{sh}$. The equation for the potential $\psi_{sh}$ is as follows
\begin{equation}\label{psish}
\frac{d^2\psi_{sh}}{dh^2}-\frac{4r}{h^3}\psi_{sh}=en_l(h)\frac{4r}{h^3},
\end{equation}
where $ n_l(h) $ is the electron column density associated with the density $n$ by the relation $ n_l = \pi nh^3 / r $ due to the dependence
of the cross-section of the magnetic tube $ s $ on the height, $ s = \pi h^3 / r $. Since the thickness of the electron layer is initially small, 
we can assume
$ n_l = N_e \delta (h-h_m) $, where $ \delta (x) $ is the Dirac delta function. The solution of equation (\ref{psish}) is
$$
\psi_{sh}=-\frac{eN_er^2}{h_m^3}\left(\frac{hh_m}{r^2}\right)^{3/4}\left\{
\begin{array}{rcl}
\hspace{-0.4cm}\exp\left(-\frac{4r^{1/2}}{h^{1/2}}+\frac{4r^{1/2}}{h_m^{1/2}}\right), \, h<h_m \\
\hspace{-0.2cm}\exp\left(-\frac{4r^{1/2}}{h_m^{1/2}}+\frac{4r^{1/2}}{h^{1/2}}\right), \, h>h_m. 
\end{array}
\right.
$$
We see that the electric field of the electron layer decreases exponentially on both sides. The characteristic width of the additional potential, $ \delta h $,
does not depend on the value of the potential, and is equal to $ \delta h \simeq h_m (h_m / r)^{1/2} $, i.e. it is equal to the width of the magnetic tube 
at the altitude $ h = h_m $. 
The total potential $ (e |\psi + \psi_{sh}| + \psi_T) / e |\psi_0| $ is shown in Figure 2 with account of the additional electric field created by 
electrons trapped into the minimum of the potential.
\begin{figure}
\begin{center}
\includegraphics[width=6cm]{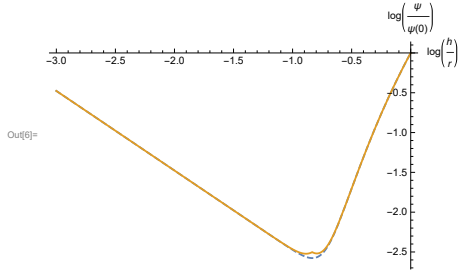}
\end{center}
\caption{The logarithms of total potential $ e|\psi + \psi_{sh}| + \psi_T $ acting on electrons (solid line) and of the potential $ e |\psi|+\psi_T $, to which electrons are trapped initially (dotted line).}
\end{figure}

\begin{figure}
\begin{center}
\includegraphics[width=6cm]{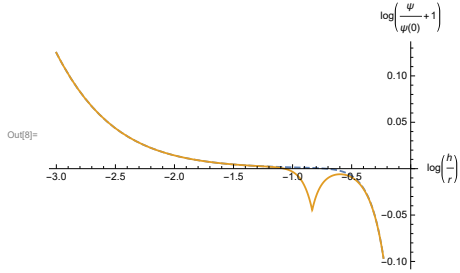}
\end{center}
\caption{The logarithm of potential $ e(\psi + \psi_{sh}) + \psi_T $ acting on positrons (solid line) and of the potential $ e\psi +\psi_T $ 
without influence of trapped electrons (dotted line).}
\end{figure}

At the height $ h \simeq h_m $ electrons form a local potential well, to which positrons are attracted. Figure 3 shows the potential
acting on positrons, $ [e (\psi + \psi_{sh}) + \psi_T] / e |\psi_0| $. In the absence of the potential of the electron layer $ \psi_{sh} $, positrons 
are accelerated by the
thermal radiation up to energy $ \gamma_0 / \kappa $, then get even more energy from the potential $ \psi $. However, if we do not
take into account the interaction of positrons passing through the layer with electrons, then positrons are not trapped inside the layer. 
Deceleration of positrons
due to bremsstrahlung radiation in the potential $ \psi_{sh} $ is not enough for them to lose energy $ \gamma_0 / \kappa $.
Here we should take into account
the collective interaction of passing positrons having the Lorentz factor $ \gamma^{+} = \gamma_0 / \kappa $ with electrons
and positrons trapped earlier. The plasma density in the layer is equal to $ n_p = n^{-} + n^{+} $, their plasma frequency is 
$ \omega_p^2 = 4 \pi n_p e^2 / m_e $.
A beam of positrons with a density $ n_b = S \lambda r / \pi c h_m^3 $ excites plasma oscillations in the layer, which decelerate positrons due to
beam instability. The increment of the beam instability $ \nu_b $ is 
$$
\nu_b=\frac{3^{1/2}}{2^{4/3}}\left(\frac{n_b}{n_p\gamma^{+3}}\right)^{1/3}\omega_p, \, n_p>n_b/\gamma^{+3}.
$$
Positrons are trapped into the well $ \psi_{sh} $, if the condition $ \nu_b \tau_p> 1 $ is satisfied, where $ \tau_p $ is the time of passage of positrons though the plasma layer, 
$ \tau_p = (h_m^3 / r)^{1/2} / c $.
Let us show for characteristic values of quantities that the condition $ \nu_b \tau_p> 1 $ 
is satisfied already at low densities of the
plasma in the layer. We take characteristic values: $ h_m \simeq 10^8 \, cm, r \simeq 10^9 \, cm, s_m = \pi h_m^3 / r \simeq 3 \cdot 10^{15} \, cm^2,
\tau_p = (h_m^3 / r)^{1/2} / c \simeq 10^{-3} \, s, \lambda \simeq 10^5, \, \gamma^{+} \simeq 3 \cdot 10^4, \, \omega_p = 5.6 \cdot 10^4 n_p^{1/2}, \, n_b = S \lambda / s_m c \simeq 2 \cdot 10^{18} \eta \, cm^{-3} $.
As a result we obtain the condition
$$
n_p>2\cdot 10^{16}\left(\frac{n_b}{1 \, cm^{-3}}\right)^{-2} \, cm^{-3},
$$
which is obviously valid for $ n_p> n_b / \gamma^{+3} \simeq 10^{5} \eta \, cm^{-3} $.
Thus, we see that
the thin layer of width $ (h_m^3 / r)^{1/2} $ arises in the polar region, where
all electrons and positrons produced in the polar magnetosphere are accumulated. Plasma is practically neutral, subrelativistic electron-positron plasma in this layer.
Electrons are confined there by the
locking potential $ \psi $ (Eq. \ref{psi}), while positrons are held by confined electrons. 

There is slight displacement of the
position of the plasma layer from the height $ h = h_m $. Indeed, the drag force of the thermal radiation of 
the star becomes twice larger than the drag force acting separately onto electrons and positrons. In the expression for the value of $ h_m $ (\ref{hm})
it is necessary to replace $ \kappa $ by $ \kappa / 2 $. As a result, the shift of the position of the plasma layer $ \Delta h_m $ is equal to
$$
\frac{\Delta h_m}{h_m}=\frac{\ln 2}{2}\left(\frac{h_m}{r}\right)^{1/2}\left[1-
\frac{5}{8}\left(\frac{h_m}{r}\right)^{1/2}\right] \simeq 0.1.
$$
\section{Burst}

Electron-positron plasma constantly accumulates in the layer. The number of pairs $ N $ grows linearly with time,
$ N = S \lambda t \simeq 2 \cdot 10^{44} \eta t $. Accordingly, the plasma column density
$ \int n_p dh = N / s_m = Nr / \pi h_m^3 \simeq 6 \cdot 10^{28} \eta t \, cm^{- 2} $ grows. The thickness relative to scattering of the
thermal radiation of the star by the
electron-positron plasma of the layer, $2 \sigma_T \int n_pdh $, grows also
with time, and becomes equal to unity at time $ t^* = s_m / 2S \lambda \sigma_T \simeq 10^{-5} \eta^{-l} \, s $.
This means that the light of the star will start to reflect effectively from the plasma layer. If the reflection coefficient
from the layer and the surface of the star is $ \eta_r $, then the energy density of the thermal radiation in the polar region under the plasma layer
will begin to grow exponentially with the characteristic time $ 2h_m / c \eta_r \simeq 10^{-2} \eta_r^{-l} \, s $. Effective
temperature of trapped radiation $ T_{eff} $ will grow in time until radiation pushes the plasma layer outside the magnetosphere.
This happens when the radiative force acting on an electron and a positron of the plasma layer,
$ 2 \sigma \sigma_T T_{eff}^4 (R / r)^2 / c $, at the height $ h = r $ becomes
equal to the force acting on electrons from the locking potential at the same height, $ e d |\psi| / dh |_{h = r} = 11e |\psi_0| / 4r $.
As a result, we have
$$
T_{eff}=10^8 K\left(\frac{11r}{8R}\kappa\right)^{1/4}\simeq 10^9 K.
$$

The released energy can be estimated as the energy of thermal radiation of the neutron star accumulated during time $ \tau_1 $, $ {\cal E}
> \pi \sigma T_{NS}^4 \tau_1 R^3 / r \simeq 10^{40} T_8^4 \, erg $.
Here we chose the minimum area of trapped radiation, $ s_R = \pi R^3 / r \simeq 3 \cdot 10^9 \, cm^2 $, which corresponds to the area of the polar region on the stellar surface.
At height $ h = h_m $ the area of trapped radiation is much larger, $ s_m \simeq 3 \cdot 10^{ 15} \, cm^2 $.
In view of the rapid growth of the energy of trapped stellar radiation, ejection of the plasma layer will occur
during a short time period $ (r-h_m) / c \simeq 3 \cdot 10^{-2} \, s $. Therefore, the power $ W $ of energy release is 
equal to $ W = c {\cal E} / r> 3 \cdot 10^{41} T_8^4 \, erg / s $.
Apparently, the radio emission mechanism here is the same as in radio pulsars - direct flux of a relativistic electron-positron
plasma is a source of radio waves. For radio pulsars an average value of the coefficient of transformation of the kinetic energy of the electron-positron
plasma into the energy of the radio emission $ \alpha_r $ is $ \alpha_r \simeq 10^{-4} $ (Beskin, Gurevich and Istomin, 1993). Thus, the energy radiated in the radio range $ {\cal E}_r $,
is equal to $ {\cal E}_r \simeq 10^{36} T_8^4 \, erg $. It is considerably smaller than the estimate given in the beginning of the paper, 
$ {\cal E}_r \simeq 10^{40} \, erg $.
The latter value was obtained under the assumption that the radiation is isotropic and has no directivity. It is clear, however, that electrons and positrons, which 
received energy from the flux of photons at the altitude $ r $ from the surface of the star, have directional motion with spread over angles 
$ \Delta \theta = R / r \simeq 10^{-3} $. Besides,
the radio emission will have the directivity $ \Delta \theta \simeq \gamma^{-1} $, where $ \gamma $ is the Lorentz factor of accelerated particles. The force
$ \sigma \sigma_T T^4 (R / r)^2 / c \simeq 1.3 \cdot 10^{-13}T_8^4 \, din $ will lead to acceleration inside the 
region $ r \simeq 10^{ 9} \, cm $ up to the energy
$ \gamma \simeq 2 \cdot 10^2T_8^4 $. This value will determine the directivity of the radio emission, $ \Delta \theta \simeq 5 \cdot 10^{-3} $. 
It should be noted that
this value of directivity follows from observations of the source FRB 121102. The observed breaks in the series of bursts have approximately
the same duration as the series themselves. If one interprets breaks as departures of the polar region from the line of sight of the observer due to rotation of the 
star with the period
$ P $, then $ P \simeq \tau_2 \simeq 20 \, d $. The radiation directivity of $ \simeq 10 \tau_1 / \tau_2 \simeq 2.5 \cdot 10^{-3} $ 
agrees well with the value of $ \gamma $, the energy of accelerated electrons and positrons produced in the polar cap of the magnetosphere of the star. 
Taking into account such directivity, the effective energy radiated in the radio range will increase 
$ 4 / \Delta \theta^2 \simeq 6 \cdot 10^5 $ times, when recalculated to the total solid angle $ 4 \pi $, which gives $ {\cal E}_r^{eff} \simeq 6 \cdot 10^{41} \, erg $. This is sufficient to place a source of
FRB at cosmological distances. As for the duration of the burst $ \tau $, it is defined by the size of the region of acceleration of the plasma
layer and its thickness $ \simeq r \simeq 10^9 \, cm $, and by the Lorentz factor of accelerated electrons and positrons, $ \tau \simeq (r / c) / {\bar \gamma} $.
For $ \gamma = 2 \cdot 10^2 $ the value of $ 1 /{\bar \gamma} = \ln(\gamma) / \gamma $ is $ \simeq 3 \cdot 10^{-2} $. Thus, the time $\tau \simeq 10^{-3} \, s $ is in agreement with the time observed in FRBs.
    
\section{Discussion}

From observations of the repeated source FRB 121102 we conclude that there exists a hierarchy of quasiperiods
$ \tau_1 \simeq 500 \, s; \tau_2 \simeq 2 \cdot 10^6 \, s \simeq 20 \, d; \tau_3 \simeq 2 \cdot 10^7 \, s \simeq  200 \, d $ in its radio emission. The time $ \tau_1 $ is
the time between consecutive bursts. Probably, because of the high power of the bursts, this is the time of accumulation of energy, which is then released during 
short time, considerably less than $ \tau_1 $. Time $ \tau_2 $ is the duration of a series of bursts and also the duration of long breaks
between bursts in the series. Finally, the time $ \tau_3 $ is the average time of recurrence of bursts. Short burst time
$ \tau \simeq 10^{- 3} \, s $ tells us about compactness of the source of radiation, apparently, a neutron star. 

Neutron star observed as radio pulsars have strong
magnetic field $ B_0 \simeq 10^{12} \, G $ and small rotation periods $ P = 10^{-3} - 1 \, s $.  The electric field, which has 
nonzero projection onto magnetic field, arises in the magnetosphere due to the rotation of a magnetized body. This occurs in the open magnetosphere,
in which magnetic field lines reach the light cylinder surface $ c/\Omega_{NS} $. Near the stellar surface the polar
cap is formed, in which continuous generation of the electron-positron plasma and the electric current takes place. For fixed value of the magnetic field
$ B_0 $, continuous plasma generation is possible for sufficiently fast rotation of the star, $ P <(B_0 / 10^{12} G)^{8/15} \, s $. The radio emission from a 
pulsar is due to the flow of relativistic plasma in the open magnetosphere (Beskin, Gurevich and Istomin, 1993). The energy of radio emission is a small fraction of the energy of the plasma flow, which
in turn is fed from the energy of rotation of the star.  Continuous plasma generation becomes impossible for slower rotation. However,
plasma generation can occur in flares, as it is in the case of RRATs. The electric field in the polar magnetosphere begins to accelerate electron-positron pairs produced by accidental 
energetic gamma quantum. This leads to a cascade plasma production and to a flash of
radio emission. For even slower stellar rotation the electric field, $ E \simeq \Omega_{NS} RB / c $, is not sufficient for cascade plasma production
in the magnetosphere. However, the external action of plasma flow, having velocity $ u $ relative to the neutron star, produces sufficiently strong
electric field. It generates cascade plasma production in the polar magnetosphere. The process of plasma  birth in this case has special
characteristics in comparison with the one, which takes place in the magnetosphere of a radio pulsar. Acceleration of electrons, injected from the plasma stream into the magnetosphere, occurs toward the star (Aurora). In this case the production of gamma rays by inverse Compton scattering of thermal photons is most efficient.
The multiplication factor $ \lambda $ (the number of pairs per one primary electron) reaches the value of $ 10^5 $. Created electron-positron plasma is
accumulated in the polar magnetosphere in the form of a thin layer in the region of a minimum of the total potential created by the electric field, which is generated
by flowing plasma, and the radiation pressure of the thermal radiation of the star (see Fig. 4). The electron-positron plasma density in the layer increases linearly
with time until the reflection of thermal photons from the layer becomes
so large that the layer becomes a mirror for the thermal radiation. After that the effective
temperature of radiation of the star under the mirror begins to grow exponentially up to the value of $ T_{eff} \simeq 10^9 \, K $. Finally, the stellar radiation
pushes the electron-positron layer outward overcoming the external blocking electric potential for a short time $ \simeq 3 \cdot 10^{-2} \, s $.  A "hole" in the magnetosphere of a neutron star
and in the stream of plasma flowing around the star forms. Accelerated up to energies $ \gamma \simeq 2 \cdot 10^2 $, the flow of electrons and positrons generates radio emission by the 
mechanism analogous to the mechanism of radio emission from pulsars. It is worth noting that the Lorentz factor of  accelerated plasma is the same as in the
secondary electron-positron plasma in the magnetosphere of a pulsar. Because the flow is relativistic, the radio emission is directed into a small solid
angle, $ 6\cdot 10^{-5} $ times smaller than $ 4 \pi $. This allows for the total energy of the burst, radiated in the radio range, $ {\cal E}_r \simeq 10^{36} \, erg $,
to appear as effective energy of isotropized burst more than five orders of magnitude higher, $ {\cal E}_r^{eff} \simeq 10^{41} \, erg $. Duration of radio emission
$ \tau $ is small, $ \tau \simeq 10^{-3} \, s$, also because of relativistic nature of the plasma flow.

It should be noted that this region of polar magnetosphere, $h\simeq h_m$, is a source of gamma radiation in the range of annihilation line $511\, KeV$ just before the burst of radio emission during the time $\tau_1$, when there is accumulation of electron-positron plasma in the layer.
The plasma density in the layer can be estimated from the condition $2n_p h_m\sigma_T\simeq 1$, when the layer becomes the mirror, $n_p\simeq 10^{16} \, cm^{-3}$. This value will be 
also the density of gamma quanta $n_\gamma = n_p$. Correspondingly, almost isotropic flux of gamma radiation from the layer in the annihilation line is, 
$S_\gamma=n_\gamma c s_m\simeq 10^{42} \, s^{-1}$,
and the luminosity in annihilation line is $\simeq 10^{36} \, erg/s$.

\begin{figure}
\begin{center}
\includegraphics[width=8cm]{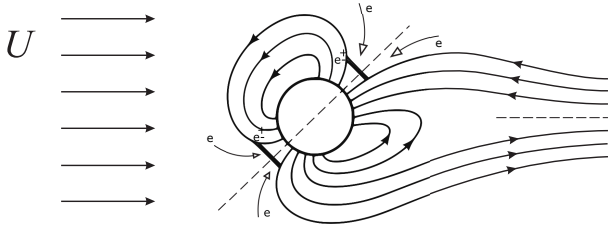}
\end{center}
\caption{Scheme illustrating flowing up of a magnetosphere of a magnetized neutron star by a plasma. Thin layers of electron-positron
plasma are formed in polar regions. Thermal radiation of the star is locked by this layers.}
\end{figure}

\section{Acknowledgments}
This work was supported by Russian Foundation for Fundamental Research, grant numbers 15-02-03063 and 16-02-00788.


\begin{thebibliography}{99}

\bibitem{blp} 
 Berestetskii V.B., Lifshitz E.M., Pitaevskii L.P., 1982, Quantum Electrodynamics, Butterworth-Heinemann

\bibitem{bgi}
 Beskin V.S., Gurevich A.V., Istomin Ya.N., 1993, Physics of the Pulsar Magnetosphere, Cambridge University Press

\bibitem{ch}
 Chatterjee S., et al., 2017, Nature, 541, 58

\bibitem{dwwh}
 Dai Z.G., Wang J.S., Wu X.F., Huang Y.F., 2016, ApJ, 829, 27

\bibitem{dd}
 Di X., Dai Z.G., 2017, ApJ, 846, 130

\bibitem{he}
 Hankins N.H., Eilek, 2007, Ap.J., 670, 693

\bibitem{is1}
 Istomin Ya.N., Sob'yanin D.N., 2011a, J. Exp. Theor. Phys., 113, 605

\bibitem{is2}
 Istomin Ya.N., Sob'yanin D.N., 2011b, Astron. Lett., 37, 468

\bibitem{k1}
 Katz J.I., 2017a, MNRAS, 467, L96

\bibitem{k2}
 Katz J.I., 2017b, MNRAS, 469, L39

\bibitem{lb}
 Lorimer D. R., Bailes M., McLaughlin M. A., Narkevic D. J., Crawford F., 2007, Science, 318, 777 

\bibitem{mz}
 Mottez F., Zarka P., 2014, A\& A, 569, A86

\bibitem{sp}
 Soglasnov V. A., Popov M. V., Bartel N., Cannon W., Novikov A. Yu., Kondratiev V. I., Altunin V. I., 2004, Ap.J., 616, 439




\end{thebibliography}
\end{document}